# An LLM-Driven Multi-Agent Debate System for Mendelian Diseases

Xinyang Zhou[1,2†], Yongyong Ren[2†], Qianqian Zhao[3†], Daoyi Huang[1], Xinbo Wang[1,2], Tingting Zhao[4], Zhixing Zhu[4], Wenyuan He[11], Shuyuan Li[7], Yan Xu[7], Yu Sun[9,10], Yongguo Yu[9,10*], Shengnan Wu[4*], Jian Wang[7,8*], Guangjun Yu[4,5,6*], Dake He[11*], Bo Ban[3*], Hui Lu[1,2,4*]

[1]Department of Bioinformatics and Biostatistics, School of Life Sciences and Biotechnology, Shanghai Jiao Tong University, Shanghai 200240, China

[2]SJTU-Yale Joint Center for Biostatistics and Data Science, Technical Center for Digital Medicine, National Center for Translational Medicine, Shanghai Jiao Tong University, Shanghai 200240, China

[3]Department of Endocrinology, Affiliated Hospital of Jining Medical University, Jining Medical University, Jining Shandong 272029, China

[4]Shanghai Children's Hospital, School of Medicine, Shanghai Jiao Tong University, Shanghai 200062, China

[5]The Second Affiliated Hospital, School of Medicine, The Chinese University of Hong Kong, Shenzhen & Longgang District People's Hospital of Shenzhen, Shenzhen 518172, China

[6]Center for Biomedical Informatics, Shanghai Engineering Research Center for Big Data in Pediatric Precision Medicine, Shanghai Children's Hospital, Shanghai 200062, China

[7]International Peace Maternity and Child Health Hospital, School of Medicine, Shanghai Jiao Tong University, Shanghai 200003, China

[8]Department of Medical Genetics and Molecular Diagnostic Laboratory, Shanghai Children's Medical Center, Shanghai Jiao Tong University School of Medicine, Shanghai 200127, China

[9]Clinical Genetics Center, Xinhua Hospital, Shanghai Jiao Tong University School of Medicine, Shanghai 200092, China

[10]Shanghai Institute for Pediatric Research, Shanghai 200092, China

[11]Department of Children Neurology, Xinhua Hospital Affiliated to Shanghai Jiao Tong University School of Medicine, Shanghai 200092, China.

† These authors contributed equally to this work.

* To whom correspondence should be addressed. Email: huilu@sjtu.edu.cn

**Abstract**

Accurate diagnosis of Mendelian diseases is crucial for precision therapy and assistance in preimplantation genetic diagnosis. However, existing methods often fall short of clinical standards or depend on extensive datasets to build pretrained machine learning models. To address this, we introduce an innovative LLM-Driven multi-agent debate system (MD2GPS) with natural language explanations of the diagnostic results. It utilizes a language model to transform results from data-driven and knowledge-driven agents into natural language, then fostering a debate between these two specialized agents. This system has been tested on 1,185 samples across four independent datasets, enhancing the TOP1 accuracy from 42.9% to 66% on average. Additionally, in a challenging cohort of 72 cases, MD2GPS identified potential pathogenic genes in 12 patients, reducing the diagnostic time by 90%. The methods within each module of this multi-agent debate system are also replaceable, facilitating its adaptation for diagnosing and researching other complex diseases.

**Introduction**

Epidemiological studies indicate that if congenital abnormalities are considered part of the genetic burden, approximately 5.3% to 7.9% of newborns globally suffer from serious genetic diseases, primarily Mendelian disorders caused by single-gene mutations[1-3]. The Online Mendelian Inheritance in Man (OMIM) shows that current genetic research has identified approximately 6,400 Mendelian diseases with unique phenotypic features associated with about 4,500 genes[4]. Despite employing sophisticated methods like phenotypic analysis, imaging, and sequencing technologies, the diagnosis rate for Mendelian diseases remains low, between 30% to 40%[5-7]. Additionally, using traditional approaches like Whole Genome Sequencing (WGS) and Whole Exome Sequencing (WES), the average diagnostic duration for Mendelian diseases is approximately two years[8]. Even after stringent filtering by bioinformatics software and clinicians, each individual still carries hundreds of potentially pathogenic or uncertain significance mutations[9]. Moreover, many rare genetic diseases share overlapping phenotypes, making the diagnostic process more complex and increasing the likelihood of misdiagnosis[10,11]. Accurate diagnosis is essential, as it plays a critical role in preimplantation genetic diagnosis (PGD) for guiding in vitro fertilization (IVF). This not only aids in the prevention of genetic disease transmission within families but also contributes to reducing the overall prevalence of these disorders in the population[12-16].

Over the past decade, a variety of computational tools have been developed to enhance the diagnosis of genetic diseases by integrating data from phenotypic observations and genetic sequencing. Tools such as Exomiser[17], LIRICAL[18], Phen-Gen[19], and diseaseGPS[20] use statistical methods to evaluate the correlation between gene variants and phenotypic manifestations, thus facilitating the prioritization of candidate pathogenic genes. The effectiveness of these tools largely depends on the availability and quality of the statistical data utilized. Similarly, tools like eXtasy[21], Xrare[22], and AMELIE[23] leverage machine learning techniques, which require extensive datasets for supervised training or fine-tuning to optimize performance. Despite these methods prioritizing candidate genes based on variant data, the lack of natural language explanations means that definitive diagnoses of genetic diseases still require doctors and scientists to conduct further validation through extensive literature reviews[20].

Debate is an essential and inevitable component of scientific discourse, serving as a critical mechanism for resolving controversies. It prompts participants to explore issues from various perspectives, which is especially important in medical diagnostics[24]. In practices like Multi-Disciplinary Treatment (MDT), debate is key to effectively treating complex medical conditions and reaching evidence-based conclusions[25,26]. However, the greatest challenge in applying debate to diagnostic algorithms lies in the semantic understanding of different methodologies and the limitations in reasoning technologies.

Recently, large language models (LLMs) have shown considerable potential to enhance debate processes, especially when supported by prompt engineering techniques[27,28]. LLMs can generate relevant insights and summarize evidence by leveraging extensive correlations within natural language. However, they also have limitations, unlike bioinformatics tools, LLMs cannot offer detailed computational biological annotations. This underscores the need for a synergistic approach, integrating the analytical capabilities of LLMs with the technical expertise of data analysis tools to fully leverage their combined strengths in scientific debates[24,29].

Here, we present MD2GPS (Medical Doctor 2 GPS), an LLM-driven multi-agent debate system designed for diagnosing Mendelian diseases. We validated MD2GPS on 1,185 samples across four independent datasets. Compared to the current state-of-the-art methods, MD2GPS achieved the highest TOP1 score in all simulated datasets. We further tested MD2GPS on an additional challenging cohort of 72 cases. MD2GPS had a consistently superior performance, which has the potential to be used as an interpretable method for Mendelian diseases diagnosis.

## Results

### Construction of MD2GPS

As depicted in Fig.1, MD2GPS is composed of three distinct agents: Data Agent, Knowledge Agent, and Debate Agent. The Data Agent processes genetic variants provided in Variant Call Format (VCF) and Human Phenotype Ontology (HPO) terms, the Data Agent annotates the impact of variants, filters out potential benign variants, and ranks the pathogenic potential of the remaining variants using an integrated bioinformatics ranking algorithm. It also explains the functions of genes and the rationale behind their pathogenic rankings in natural language. Concurrently, the Knowledge Agent elucidates genes and their physiological functions, employing the embedded GPT-4.0 large language model to rank gene pathogenicity. The Debate Agent then assesses the consistency and accuracy of the evidence provided by the previous two agents, ultimately delivering the final rankings and their justifications.

After the construction of MD2GPS, we performed a head-to-head comparison of MD2GPS with four other state-of-the-art methods (diseaseGPS, Xrare, Exomiser, and LIRICAL) on four different datasets (Table 1). We considered four evaluation metrics and performed the assessments on 1185 genetic disease patients from 4 different datasets (Table 1). The evaluation metrics (TOP1, TOP3, and TOP5) measure the cumulative percentage of cases where the TOP-ranked gene is the correct pathogenic gene. This set of metrics provides a gradient measure of the diagnostic tool's ability to identify the correct gene within a broader range of TOP candidates.

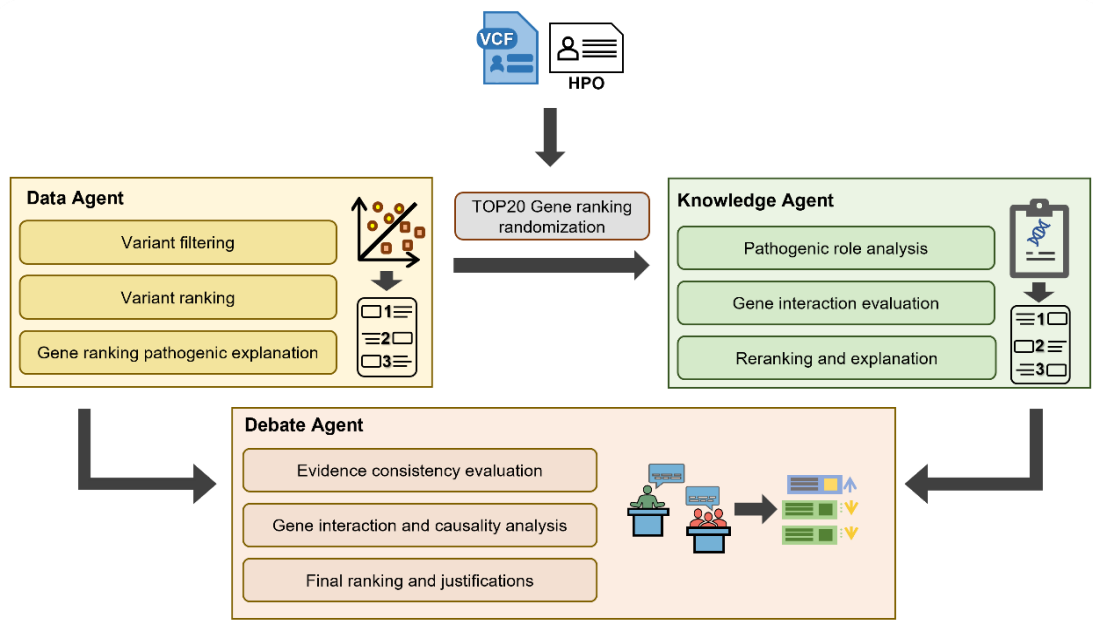

**Fig. 1: Workflow of MD2GPS.** MD2GPS integrates three functional agents to process genetic information. The Data Agent utilizes genetic data, annotating impacts, filtering variants, and ranking pathogenic potentials with an algorithm while also providing natural language explanations of gene functions. Simultaneously, the Knowledge Agent uses a GPT-4.0 model to further analyze and rank gene pathogenicity based on physiological functions. Finally, the Debate Agent evaluates the output from the other agents for consistency and accuracy, culminating in the provision of final, justified gene rankings.

**Performance comparison of MD2GPS and the current state-of-the-art methods**

Fig. 2 summarized the performance of different methods when applied to diagnosis Mendelian genetic disease. As it can been seen, MD2GPS archived the consistently highest TOP1 score in all datasets, which scored the average TOP1 of 66.0%, while the rest of the methods scored lower as follows: 47.0% (Xrare), 42.9% (diseaseGPS), 34.0% (LIRICAL), and 33.1% (Exomiser). Additionally, in the TOP3 and TOP5 rankings, MD2GPS also demonstrates superior performance. When evaluated using the TOP5 metric, MD2GPS increased the cumulative percentage of cases where the TOP 5 genes include the correct pathogenic gene from 55.2%-73.2% to 85.1%.

MD2GPS consistently outperformed other methods, achieving the highest TOP1 scores across all datasets, including both simulated and real-case scenarios involving single or multiple phenotypes. In the JN dataset, which focuses on a single phenotype of Dwarfism, MD2GPS achieved a TOP1 score of 70.0%, exceeding the performance of the next best method, Xrare, by 13.3% (Fig. 2A). In an analysis of 734 patients diagnosed with developmental disorders, MD2GPS again achieved the highest TOP1 score, surpassing the next best method, diseaseGPS, by 8.2% (Fig. 2B). In scenarios involving multiple abnormal phenotypes, MD2GPS further demonstrated its superiority. In a simulated dataset (RD dataset) characterized by clear etiology, diagnosis, and treatment plans, it improved the TOP1 accuracy from 65.6% (achieved by LIRICAL) to 76.8% (Fig. 2C). In the real-world SCH dataset, a multi-expert consultation dataset, MD2GPS enhanced TOP1 accuracy from 39.6% to 65.8%, an increase of 26.2% (Fig. 2D).

We further investigated the diagnostic performance of MD2GPS and other methods within the RD dataset, specifically across diseases with known incidence rates. The diseases were sorted by their median incidence rates into four distinct quartiles: <1/125000, [1/125000, 1/75000], [1/75000, 1/15000] and >1/15000, MD2GPS consistently demonstrated the best TOP1 accuracy across all groups. As shown in Fig. 2E, in the lowest prevalence quartile (<1/125000), MD2GPS achieved a remarkable 61.5% TOP1 accuracy, significantly outperforming other methods. Its performance peaked in the next quartile [1/125000, 1/75000] with an impressive 82.9% TOP1 accuracy, markedly higher than other methods. For diseases in the intermediate prevalence group [1/75000, 1/15000], MD2GPS achieved a TOP1 accuracy of 75.0%, demonstrating the best performance in this category. In the highest prevalence category (>1/15000), MD2GPS continued to lead with 73.3% accuracy, showcasing its effectiveness across diverse genetic disease prevalence.

**Table 1. Mendelian genetic disease datasets.**

| Dataset name | Sources | Dataset Description | Number of cases |
|---|---|---|---|
| DDD dataset | Deciphering Developmental Disorders (DDD) Project | We deduplicated pathogenic genes and randomly selected one case per gene in DDD project for case simulation. | 734 |
| RD dataset | Manually constructed from PubMed | The dataset covers 96 genetic diseases listed in 'China's first list of rare diseases'. | 234 |
| SCH dataset | Shanghai Children's Hospital | All cases involve complex, multidisciplinary treatment scenarios that are challenging to diagnose from real world. | 187 |
| JN dataset | Affiliated Hospital of Jining Medical University | A real single-disease dwarfism dataset. | 30 |

The SCH and RD datasets involve multiple genetic diseases; SCH includes complex, multidisciplinary treatment scenarios, RD covers 95 genetic diseases listed in 'China's first list of rare diseases'. The JN and DDD datasets focus on specific disease presentations: JN features cases of dwarfism, DDD includes unique pathogenic genes deduplicated from cases of Deciphering Developmental Disorders.

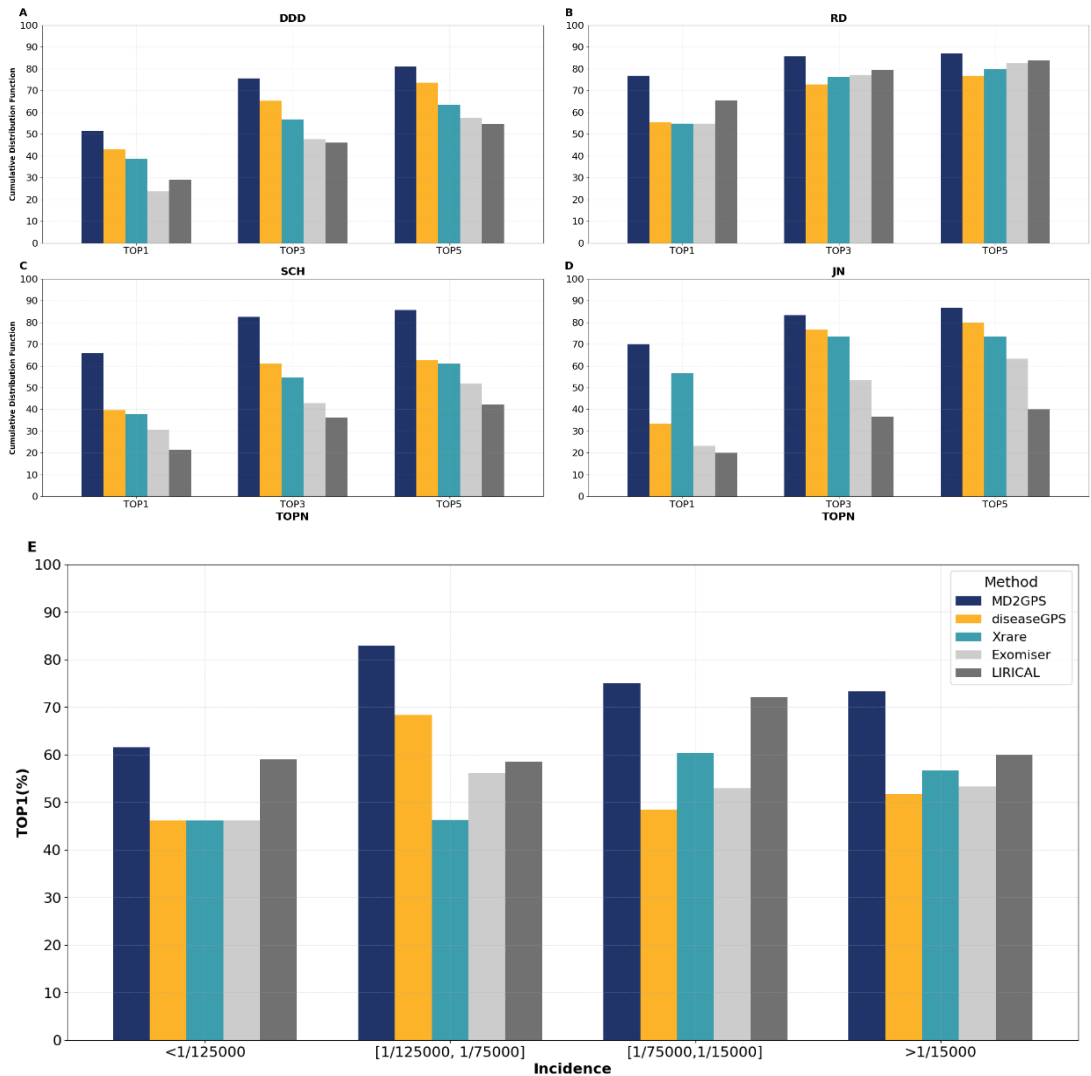

**Fig. 2: The performance of diagnosis methods when applied to four Mendelian genetic disease datasets.** Performance evaluation of all methods was conducted using the Cumulative Distribution Function (CDF) across two simulated datasets, A) DDD and B) RD, and two real datasets, C) SCH and D) JN. E) The figure summarizes the comparative performance of the MD2GPS and other diagnostic tools on the RD dataset, displaying the TOP1 accuracy across different incidence rate categories: <1/25000, [1/25000, 1/75000], [1/75000, 1/15000], and >1/15000.

**Evaluating the Role of the Debate Agent in Enhancing Diagnostic Accuracy**

To evaluate the impact of the Debate Agent on model performance, we performed a comparative analysis between the outputs of the Knowledge Agent and the Debate Agent. Both agents utilize the advanced GPT-4 large language model to rank gene pathogenicity. The analysis indicates that the diagnostic performance significantly improves after the debate, with the TOP1 accuracy increasing by 13.4% compared to the average of the Knowledge Agent. This improvement was particularly significant in real-world complex cases, as evidenced in the SCH dataset, where the increase in accuracy reached 16.6%. These findings highlight the essential role of the Debate Agent in enhancing diagnostic accuracy by more effectively assessing the consistency and relevance of evidence.

We compared the diagnostic results of a single LLM-driven Knowledge Agent with those obtained after a multi-agent debate. The inclusion of the Data Agent brought diverse perspectives essential for deeper debates, while the introduction of the Debate Agent facilitated strategic enhancements. Extensive testing across several datasets revealed that a multi-agent debate significantly bolsters LLMs' diagnostic capabilities for genetic disorders (Fig. 3). Remarkably, the model's accuracy for the most critical predictions (TOP1) increased by 10% to 16.6%, while stability was maintained in the TOP5 predictions. These results affirm the efficacy of the multi-agent debate strategy and highlight its essential contribution to enhancing the model's precision in identifying key diagnostic outcomes.

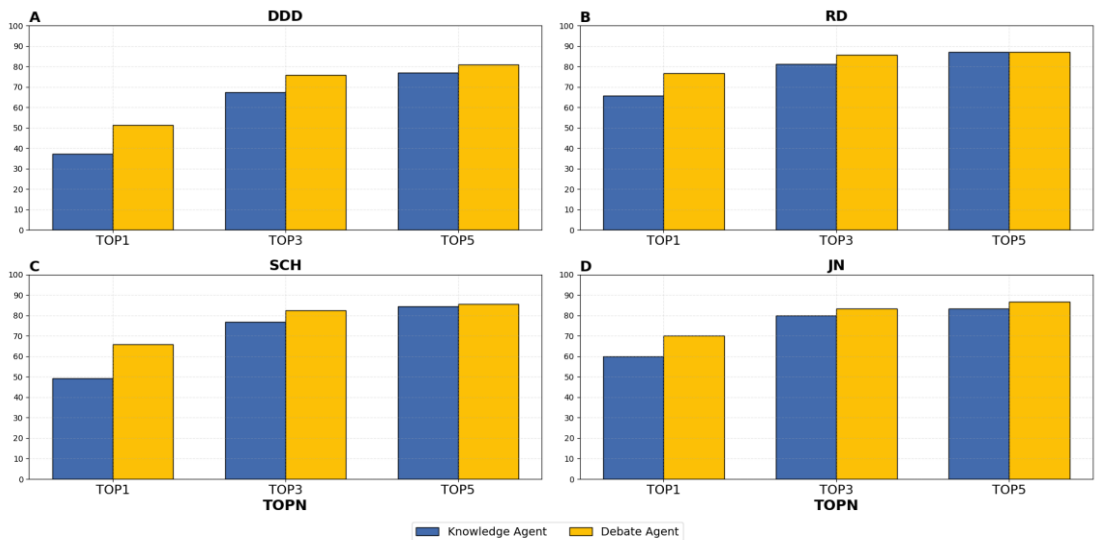

**Fig. 3: The performance of Knowledge Agent and Debate Agent when applied to four Mendelian genetic disease datasets.** This figure shows the diagnostic accuracy of Knowledge Agent (yellow bars) and Debate Agent (blue bars) across four datasets labeled A (DDD, synthetic), B (RD, synthetic), C (SCH, real), and D (JN, real).

To better understand how debates impact performance, we categorized agent responses into eight distinct patterns based on whether the TOP1 answer identified the pathogenic gene. Each pattern is denoted by a sequence of responses from the Knowledge, Data, and Debate Agents, where 'A' indicates that the pathogenic gene is ranked first and 'B' signifies it is not. The categorization includes consistent answers (where all three agents agree, marked as A/A/A or B/B/B), recall answers (where the Knowledge or Data Agents do not identify the pathogen as TOP1, but the Debate Agent does, as in B/B/A, A/B/A, and B/A/A), and missing answers (where at least one of the Knowledge or Data Agents ranks the pathogen as TOP1, but the Debate Agent does not, such as B/A/B, A/B/B, and A/A/B).

This structured analysis offered critical insights into the dynamics of the diagnostic process. As shown in Fig. 4A, the statistical findings indicate that consistency rates across all datasets exceeded 60%, with the RD dataset achieving the highest accuracy at 75% and the DDD dataset the lowest at 62.4%. The high consistency rates suggest strong agreement among the agents, particularly in B/B/B scenarios, where none of the agents ranked the pathogenic gene as TOP1, highlighting a unified diagnostic perspective despite the negative outcome.

Furthermore, in the DDD, RD, and SCH datasets, recall rates consistently exceeded missing rates, suggesting that the Debate Agent effectively enhanced diagnostic accuracy by recalibrating initial discrepancies. The JN dataset exhibited the highest recall rate at 30%, while maintaining a missing rate of 0%, indicating that multi-agent debates remain effective even in scenarios with fewer observed phenotypes, such as Dwarfism (Fig. 4). These findings underscore the pivotal role of the Debate Agent in not only confirming correct initial diagnoses but also effectively rectifying erroneous ones. This effectiveness is evident in the other datasets, where the Debate Agent successfully corrected inaccuracies made by both the Knowledge and Data Agents (category B/B/A). This capability to dynamically refine diagnostic assessments through structured debates enhances the potential of multi-agent systems to achieve higher diagnostic accuracies and discover new medical insights beyond existing knowledge.

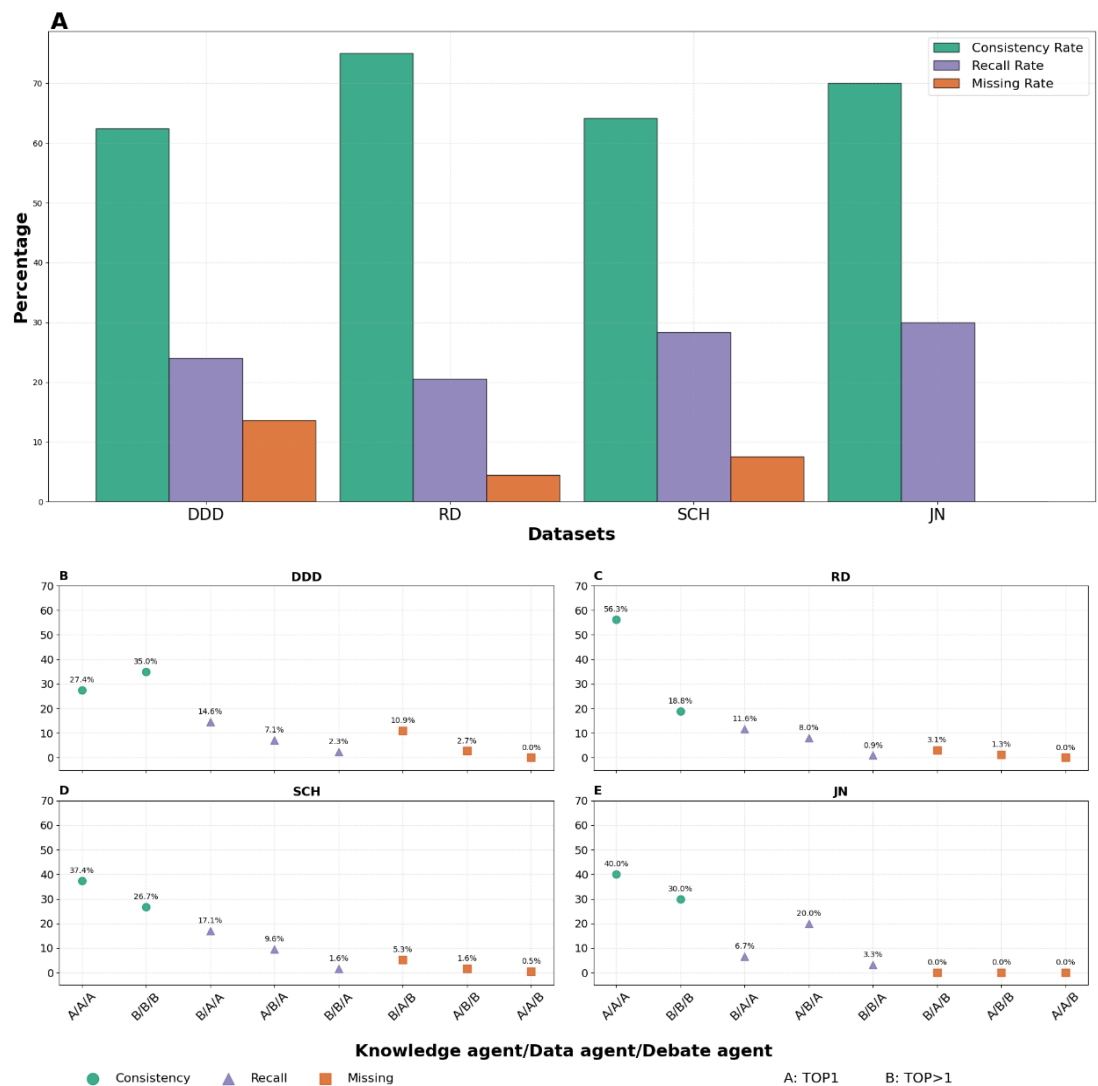

**Fig. 4: The Performance of MD2GPS Response Patterns Across Four Datasets.** A) Summary of consistency, recall, and missing rates across four datasets. BCDE) The statistical analysis categorizes multi-agent responses into eight groups, each containing sequential responses from the Knowledge, Data, and Debate Agents. In this system, 'A' indicates the pathogenic gene is ranked first, while 'B' signifies it is no. The green circles represent consistent answers, the purple triangles indicate recalled answers, and the orange rectangles signify missing answers.

To further demonstrate the functionality of MD2GPS, we discuss a case of a B/B/A. In this example, the Knowledge Agent, Data Agent, and Debate Agent each propose PEX6, HLA-DRB1, and TBC1D24 as pathogenic genes, respectively. They evaluate the patient's phenotypes from their unique analytical perspectives.

Specifically, the Knowledge Agent emphasizes hearing impairment and hand abnormalities as pivotal phenotypes. In contrast, the Data Agent, through its analytical computations, prioritizes fever as the most significant phenotype. After a comprehensive debate, the Debate Agent adjusts the earlier evaluations and determines epilepsy to be the patient's primary and most severe phenotype. This leads to the recognition that TBC1D24, initially ranked highest by the Debate Agent, is the gene most closely associated with the neurological disorders presented by the patient. This case highlights MD2GPS's ability to integrate varied diagnostic perspectives and refine gene-phenotype correlations through its multi-agent debate system. This case exemplifies the nuanced decision-making process facilitated by the MD2GPS system, showcasing its ability to integrate diverse diagnostic insights and refine the final gene-phenotype associations through debate.

**Assessing the Extended Performance and Clinical Potential of MD2GPS on Additional Real-World Datasets**

We conducted a prospective clinical application of the MD2GPS system at the Affiliated Hospital of Jining Medical College, involving 320 patients diagnosed with dwarfism. Initially, physicians applied ACMG standards to establish evidence levels for identified mutations, confirming 72 as pathogenic or likely pathogenic. The debate function of MD2GPS then facilitated further analysis, which, alongside literature consultation, led to the identification of causative genes in 12 patients. Remarkably, the entire diagnostic process for the cohort of 320 patients was completed in just 11 days, marking a 90% reduction in time compared to the initial four-month duration required for their first diagnosis in three years ago. Within these cases, MD2GPS achieved TOP1 and TOP5 accuracies of 66.7% and 100%, respectively, significantly outperforming other systems like Xrare and diseaseGPS.

The study also uncovered discrepancies in 60 cases where identified mutations did not align with observed phenotypes, suggesting potential polygenic interactions that require further exploration. These findings not only demonstrate MD2GPS's efficiency and accuracy in a clinical setting but also its capacity to unearth new avenues for medical research, underlining the system's substantial potential to expand existing medical knowledge.

**Discussion**

In this study, we introduced MD2GPS, an innovative LLM-driven multi-agent debate system aimed at enhancing the reliability and efficiency of genetic disease diagnostics. Diverging from traditional models like Xrare and AMELIE, which require extensive training and may suffer from overfitting with fewer rare genetic disease samples, MD2GPS operates without specific training requirements and has demonstrated a 23.1% improvement from 42.9% to 66% on average in the TOP1 score. In clinical settings, MD2GPS has not only refined the precision of diagnoses but also significantly reduced the diagnostic timeframe by approximately 90%, markedly streamlining the genetic diagnostics process.

To accommodate continual technological advancements, MD2GPS was designed with a modular architecture, allowing for easy updates to critical components. This design permits seamless replacement of the ranking algorithms in the Data Agent, as well as the LLMs and prompts in the Knowledge and Debate Agents. Such flexibility ensures that MD2GPS remains at the forefront of genetic diagnostics, readily adapting to emerging technologies.

MD2GPS was developed and assessed primarily for Mendelian disease samples involving SNV and Indel variants, where it exhibited excellent performance. However, MD2GPS is also capable of handling other types of genetic variants, including non-coding variants, structural variations, and copy number variations (CNVs). It allows for the integration of these novel variant types into the Data Agent, enabling users to either input their results directly or incorporate their own bioinformatics pipelines. The primary limitation of MD2GPS is that the Data Agent currently supports only single-sample analysis, as opposed to family-based sample analysis. Nevertheless, users can still input results from family sample analyses into the Data Agent for subsequent debate and further analysis.

Furthermore, although MD2GPS is designed for diagnosing genetic diseases, the multi-agent debate framework is also applicable to the diagnosis of other complex conditions, such as

Alzheimer's and Parkinson's disease. These neurodegenerative diseases are multifactorial, involving complex interactions between genetic predispositions and environmental factors. By incorporating algorithms that identify genetic markers associated with these conditions into the Data Agent, MD2GPS can provide a preliminary genetic assessment. Additionally, the Knowledge Agent can be customized to include prompts based on the latest research into lifestyle and environmental factors that influence disease progression. During debates, the agents can assess and discuss the significance and impact of various genetic and non-genetic factors, offering a comprehensive view that integrates both current analyses and existing knowledge.

MD2GPS not only ranks potential pathogenic genes but also elucidates the reasons for their ranking, explaining why a particular genetic variant is ranked at the top and not others. This level of interpretability is crucial for analyzing multi-genic effects, offering significant value in understanding complex genetic interactions. This was demonstrated in the case of a six-year-old diagnosed with Noonan Syndrome, where MD2GPS aligned with the clinical top diagnosis of a KRAS p.V14I mutation and also highlighted an MTOR mutation as equally pathogenic. This insight suggested a polygenic effect, potentially exacerbating the patient's symptoms such as neck webbing—a phenotype not typically associated with the less harmful KRAS p.V14I mutation alone. By revealing such intricate genetic relationships that might be overlooked by traditional models, MD2GPS provides deeper insights and aids in developing more targeted therapeutic strategies.

In conclusion, MD2GPS revolutionizes Mendelian disease diagnosis through its LLM-driven, debate-centric approach. By leveraging a multi-agent system, it not only outperforms existing diagnostic methods but also offers a more dynamic and expandable platform for the continual improvement and application of genetic diagnostics. This system not only enhances the accuracy and efficiency of diagnosing genetic disorders but also plays a crucial role in advancing our understanding of genetic pathologies and their treatment options.

## Online Methods

### Data preparation

This study utilized 1,257 genetic disease samples from four distinct datasets: two real datasets from Shanghai Children's Hospital (defined as the SCH dataset) and Jining Medical University

Hospital (defined as the JN dataset), and two constructed datasets based on public genetic disease information from the DDD database (defined as the DDD dataset) or case reports in the literature (defined as the RD dataset).

The SCH dataset includes 187 patients diagnosed with genetic diseases at Shanghai Children's Hospital, which was used in our previously study[20]. It encompasses 326 disease categories, over 380 abnormal phenotypes, and 140 different pathogenic genes identified as causative. All patients were challenging diagnoses for primary care physicians and were ultimately diagnosed through consultations by a multi-disciplinary team.

The JN dataset comprises 102 patients from the Affiliated Hospital of Jining Medical University, which has not been previously proposed. It includes patients primarily presenting with short stature, with no significant abnormalities in physical signs or physiological and biochemical tests noted. Of these, 30 cases were confirmed with identifiable causative genes, involving 12 abnormal phenotypes and 20 different causative pathogenic genes. The remaining 72 samples, identified as harboring pathogenic or likely pathogenic variants according to ACMG guidelines yet lacking conclusively identified causative genes, exhibit a consistent set of clinical phenotypes: Pituitary dwarfism (HP:0000839), Intrauterine growth retardation (HP:0001511), Severe short stature (HP:0003510), and Short stature (HP:0004322). Beyond these, no additional specific abnormal phenotypes have been observed.

The DDD dataset consists of phenotypic and genotypic data from the Deciphering Developmental Disorders project[30]. Initially, 6,085 cases were screened through the OMIM database[4], identifying those with definitive genetic disorders based on the causal genes and their modes of inheritance. Subsequently, to ensure diversity and non-redundancy, we randomly selected one sample from each group of cases with identical causal genes, resulting in a subset of 734 samples. The variant loci within each case were then randomly assigned to one of 157 healthy individual genomes, which served as the background genomes. These 157 healthy genomes, provided by Shanghai Children's Hospital, were independent and unrelated to the DDD study. This dataset encompasses 1242 diseases, over 1390 abnormal phenotypes, and 734 distinct causal pathogenic genes.

The RD dataset was developed based on China's first official list of rare diseases, which includes 121 distinct diseases, each with an established diagnostic and treatment protocol[31]. Of these, 26 diseases were excluded based on specific criteria: absence of sufficient case study literature, conditions caused by somatic or mitochondrial DNA variations. The remaining 95 diseases, each represented by two or three cases, contributed to a total of 234 samples. Phenotypic data and variant descriptions were manually extracted and summarized from public literature sources. Genetic information, including gene names, cDNA mutations, and the corresponding protein changes, were used to convert to genomic DNA (gDNA) coordinates using TransVar[32], with left aligned standardization on the chromosome. The raw descriptions in natural language were initially converted into HPO terms using PhenoGPT[33] and then meticulously reviewed and refined by two experienced experts. Consequently, the RD dataset now contains 95 diseases, over 870 abnormal phenotypes, and 144 different causal pathogenic genes.

**Data Agent**

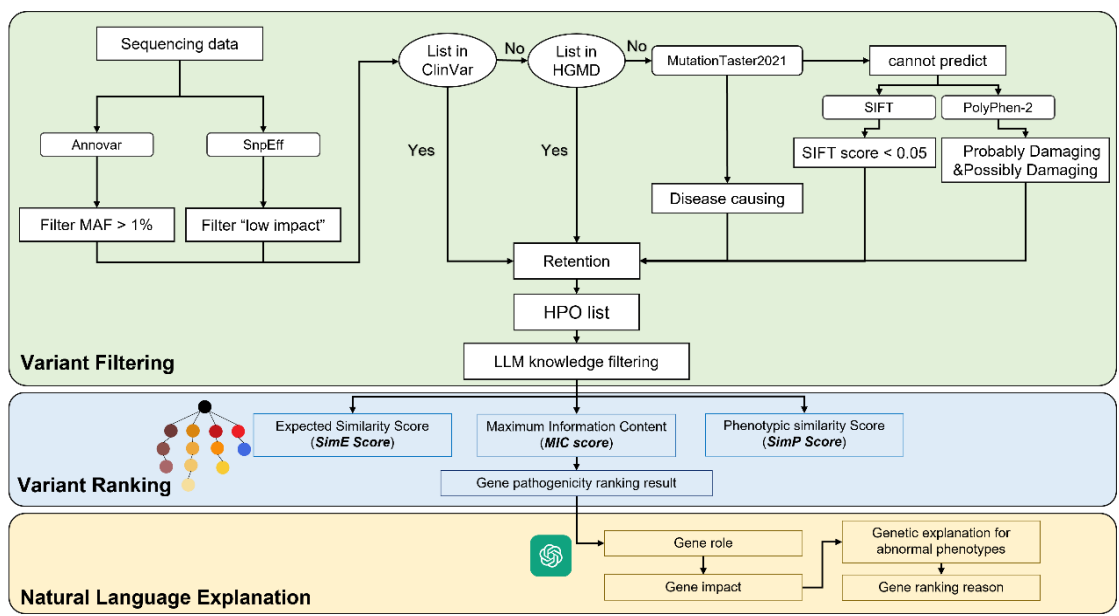

**Fig. 5: Functional architecture of the Data Agent integrating bioinformatics data analysis with LLM explanation.** The Data Agent integrates three modules: variant filtering, variant ranking, and natural language explanation. Gene data, refined through bioinformatics and phenotype-based GPT-4 filters, are ranked using an integrated algorithm. GPT-4 then generates structured natural language interpretations of these ranked results.

The Data Agent within MD2GPS is comprised of three main modules: variant filtering, variant ranking, and natural language explanation (Fig. 5). In the variant filtering process, the genetic variants in variant call format (VCF) are annotated using ANNOVAR[34] and snpEff[35], which provide comprehensive details such as gene location, amino acid changes, allele frequency in the population, data-based predictions of variant impact, and annotations from databases like Human Gene Mutation Database (HGMD)[36] and ClinVar[37]. Then, the annotated variants are filtered in stages, starting with the exclusion of variants exhibiting a minor allele frequency (MAF) greater than 1% in the general population. Subsequently, variants classified as "Benign" or "Likely Benign" in the ClinVar database are filtered out. Variants not covered by ClinVar or marked as "Uncertain" are retained if they are documented as pathogenic in the HGMD. Further refinement is applied by discarding variants predicted as 'polymorphism' by MutationTaster2021[38]. For those not assessable by MutationTaster2021, those with a SIFT score[39] of ≥0.05 and a non-pathogenic prediction from PolyPhen-2[40] are eliminated. Finally, the integrated GPT-4 model is employed to evaluate the relevance of each gene's functional impairment to at least one of the patient's phenotypes, discarding any gene completely unrelated to the phenotypes.

In the variant ranking process, three key metrics are calculated to score the variants: the Disease Empirical Similarity Score (*SimE*), the Maximum Information Content of Clinical Phenotypes (*MIC*), and the Theoretical Similarity Score of a Disease (*SimT*), which aggregates phenotype-based and genotype-based disease similarity scores. Based on these metrics, all potential genetic diseases are ranked in descending order of priority, facilitating a structured approach to identify the most likely genetic conditions affecting the patient.

Lastly, the Data Agent utilizes GPT-4 to generate natural language explanations based on the ranking outcomes. This involves articulating the functions of genes, detailing the molecular mechanisms linking genes to diseases, and explaining the reasons behind the pathogenicity rankings of genes associated with the diseases. This narrative layer not only aids in understanding the computational results but also enhances the communication of findings to clinical practitioners and researchers, ensuring clarity and actionable insights.

**Knowledge Agent**

Within the MD2GPS framework, the Knowledge Agent employs the GPT-4 model to process unordered mutation filtering and ranking results provided by the Data Agent. This allows for comprehensive analysis of gene mutations' contributions to diseases, taking inputs such as lists of symptoms and gene mutations presented in natural language.

During the analysis, each gene mutation's role, impact on patient physiology, and ranking rationale are meticulously explored to clarify their pathogenic roles in diseases. Special emphasis is placed on the logic behind selecting specific gene sequences—not only based on each gene's biological functions and contributions to symptoms but also considering gene interactions. This thorough analysis, contrasting with other potential sequences, explains how the chosen sequences more accurately reflect the patient's actual condition.

The outcomes of Knowledge Agent are delivered in two parts: the rankings are outputted in JSON format to ensure data transparency and facilitate subsequent analysis, while the reasoning explanations are provided in natural language, offering a scientific and logical foundation for the diagnostic process.

**Multi-agent Debate**

The Debate Agent in MD2GPS debates by integrating responses from the Knowledge and Data agents, which include variant rankings, interpretations, and evidence from databases like ClinVar. Using GPT-4, the agent processes these inputs through structured prompts that guide the debate on gene pathogenicity. Inputs comprises a list of pathogenic genes that has been deduplicated and randomized based on the top 20 genes identified by the Bioinformatic ranking algorithm, formatted to address limitations associated with the GPT model's context length. Each prompt directs GPT-4 to evaluate the systemic effects of gene combinations, beyond the pathogenicity of individual genes.

During the implementation of this debate, the Debate Agent evaluates the consistency and accuracy of the evidence provided by the Knowledge Agent and Data Agent, ensuring that the information relied upon aligns with current genetic knowledge. Additionally, it conducts rigorous assessments of severity, analyzing both the immediate and long-term impacts of each phenotype, and identifies phenotypes directly caused by genetic variations to determine which gene mutations are most critical to disease progression. In the gene ranking process, priority is given to genes that influence patient phenotypes and are classified as high-confidence variations, ensuring that the focus is on the genes most likely to cause disease, thereby enhancing the specificity and efficiency of diagnostics.

Furthermore, the Debate Agent emphasizes the clarity of causal relationships, requiring a clear delineation of direct causal pathways when inferring the impact of gene mutations on diseases.

This clarity in causal relationships aids in a deeper understanding of disease mechanisms and supports the development of more effective treatment strategies. Considering the polygenic nature of diseases, the Debate Agent also comprehensively assesses the interactions between genes and their collective contributions to disease phenotypes. This comprehensive analysis reflects the complexity of diseases and necessitates a multidimensional approach to interpreting genetic data, thereby enhancing the overall understanding of the genetic background and progression of diseases. Overall, the processes are formulated as follows:

*Data Agent:*

$$\{Gene_1, Gene_2, \cdots\} \in \mathbb{S}_{gene}, \{HPO_1, HPO_2, \cdots\} \in \mathbb{S}_{HPO}, \mathbb{R}_{Gene\ ranks} \xrightarrow{LLM} [\mathbb{R}_{Data\ Agent} | \mathbb{Z}_{Data\ Analysis}] \quad (1)$$

*Knowledge Agent:*

$$\{Gene_1, Gene_2, \cdots\} \in \mathbb{S}_{gene}, \{HPO_1, HPO_2, \cdots\} \in \mathbb{S}_{HPO} \xrightarrow{LLM} [\mathbb{R}_{Knowledge\ Agent} | \mathbb{Z}_{Knowledge\ Analysis}] \quad (2)$$

*Debate Agent:*

$$\mathbb{S}_{gene}, \mathbb{S}_{HPO}, \mathbb{Z}_{Knowledge\ Analysis}, \mathbb{Z}_{Data\ Analysis}, \mathbb{S}_{ClinVar} \xrightarrow{LLM} [\mathbb{R}_{Debate\ Agent} | \mathbb{Z}_{Debate\ Analysis}] \quad (3)$$

$\mathbb{S}_{gene}$ and $\mathbb{S}_{HPO}$ in the patient's medical records represent lists of candidate genes and sets of abnormal phenotypes, respectively. The Knowledge Agent performs the analysis ($\mathbb{Z}_{Knowledge\ Analysis}$) based on the medical data, while the Data Agent utilizes both the medical data and its own calculated gene ranks ($\mathbb{R}_{Gene\ ranks}$) for analysis ($\mathbb{Z}_{Data\ Analysis}$), both agents being driven by LLMs. Subsequently, the Debate Agent integrates the analyses generated by the two agents with pathogenic evidence ClinVar ($\mathbb{S}_{ClinVar}$) provided from Data Agent, conducting a debate that results in the derivation of $\mathbb{Z}_{Debate\ Analysis}$.

**Evaluation**

In evaluating the efficacy of the MD2GPS framework for variant prioritization, a systematic approach was employed using TOPN cumulative distribution function (CDF) metrics—TOP1, TOP3 and TOP5. These metrics assess the likelihood that the causative gene ranks within the top N predictions, reflecting the cumulative percentage of cases where the causative gene is accurately identified up to each specific rank, ensuring a fair and comprehensive comparison of algorithm performance. 'Consistent answers' are defined as instances where all agents—Knowledge, Data, and Debate—uniformly agree on the gene's pathogenic status, either ranking it as TOP1 or not at all. 'Recall answers' occur when the causative gene ascends to TOP1 following debate deliberations despite initial discrepancies, and 'Missing answers' are noted when a gene drops from TOP1 post-debate. 'Consistency rates' are measured by the percentage of consistent answers, 'missing rates' by the frequency of missing answers, and 'recall rates' are calculated as the proportion of recall answers to total cases, highlighting the system's capability to rectify initial inaccuracies. Diagnostic times were specifically monitored in a cohort diagnosed with dwarfism at Jining Medical University Affiliated Hospital to assess the practical impact and speed of the diagnostic process in a real-world clinical setting.

## References

1. Church, G. Compelling Reasons for Repairing Human Germlines. *New England Journal of Medicine* **377**, 1909-1911 (2017).
2. Posey, J.E.*, et al.* Insights into genetics, human biology and disease gleaned from family based genomic studies. *Genetics in Medicine* **21**, 798-812 (2019).
3. Rahit, K.T.H. & Tarailo-Graovac, M. Genetic modifiers and rare mendelian disease. *Genes* **11**, 239 (2020).
4. Amberger, J.S., Bocchini, C.A., Scott, A.F. & Hamosh, A. OMIM.org: leveraging knowledge across phenotype-gene relationships. *Nucleic Acids Res* **47**, D1038-d1043 (2019).
5. Boycott, K.M., Vanstone, M.R., Bulman, D.E. & MacKenzie, A.E. Rare-disease genetics in the era of next-generation sequencing: discovery to translation. *Nat Rev Genet* **14**, 681-691 (2013).
6. Yang, Y.*, et al.* Clinical whole-exome sequencing for the diagnosis of mendelian disorders. *N Engl J Med* **369**, 1502-1511 (2013).
7. Trujillano, D.*, et al.* Clinical exome sequencing: results from 2819 samples reflecting 1000 families. *Eur J Hum Genet* **25**, 176-182 (2017).
8. Ewans, L.J.*, et al.* Whole exome and genome sequencing in mendelian disorders: a diagnostic and health economic analysis. *Eur J Hum Genet* **30**, 1121-1131 (2022).
9. Lek, M.*, et al.* Analysis of protein-coding genetic variation in 60,706 humans. *Nature* **536**, 285-291 (2016).
10. Stoller, J.K. The challenge of rare diseases. *Chest* **153**, 1309-1314 (2018).
11. Faviez, C.*, et al.* Diagnosis support systems for rare diseases: a scoping review. *Orphanet Journal of Rare Diseases* **15**(2020).
12. Kay, V.J. & Irvine, D.S. Successful in-vitro fertilization pregnancy with spermatozoa from a patient with Kartagener's syndrome: Case Report. *Human Reproduction* **15**, 135-138 (2000).
13. Khalifa, E.*, et al.* Successful fertilization and pregnancy outcome in in-vitro fertilization using cryopreserved/thawed spermatozoa from patients with malignant diseases. *Hum Reprod* **7**, 105-108 (1992).
14. Pirtea, P.*, et al.* Successful ART outcome in a woman with McCune-Albright syndrome: a case report and literature review. *J Assist Reprod Genet* **40**, 1669-1675 (2023).

15. Winkler, I., *et al.* A Successful New Case of Twin Pregnancy in a Patient with Swyer Syndrome-An Up-to-Date Review on the Incidence and Outcome of Twin/Multiple Gestations in the Pure 46,XY Gonadal Dysgenesis. *Int J Environ Res Public Health* **19**(2022).
16. Yu, Y., *et al.* Confrontment and solution to gonadotropin resistance and low oocyte retrieval in in vitro fertilization for type I BPES: a case series with review of literature. *J Ovarian Res* **14**, 143 (2021).
17. Smedley, D., *et al.* Next-generation diagnostics and disease-gene discovery with the Exomiser. *Nat Protoc* **10**, 2004-2015 (2015).
18. Robinson, P.N., *et al.* Interpretable Clinical Genomics with a Likelihood Ratio Paradigm. *Am J Hum Genet* **107**, 403-417 (2020).
19. Javed, A., Agrawal, S. & Ng, P.C. Phen-Gen: combining phenotype and genotype to analyze rare disorders. *Nat Methods* **11**, 935-937 (2014).
20. Huang, D., *et al.* diseaseGPS: auxiliary diagnostic system for genetic disorders based on genotype and phenotype. *Bioinformatics* **39**(2023).
21. Sifrim, A., *et al.* eXtasy: variant prioritization by genomic data fusion. *Nat Methods* **10**, 1083-1084 (2013).
22. Li, Q., Zhao, K., Bustamante, C.D., Ma, X. & Wong, W.H. Xrare: a machine learning method jointly modeling phenotypes and genetic evidence for rare disease diagnosis. *Genet Med* **21**, 2126-2134 (2019).
23. Birgmeier, J., *et al.* AMELIE speeds Mendelian diagnosis by matching patient phenotype and genotype to primary literature. *Sci Transl Med* **12**(2020).
24. Fouad, Y., *et al.* The NAFLD-MAFLD debate: eminence vs evidence. *Liver International* **41**, 255-260 (2021).
25. Garrett, M., Schoener, L. & Hood, L. Debate: A teaching strategy to improve verbal communication and critical-thinking skills. *Nurse educator* **21**, 37-40 (1996).
26. Rafailov, R., *et al.* Direct preference optimization: Your language model is secretly a reward model. *Advances in Neural Information Processing Systems* **36**(2024).
27. Lin, Z., Niu, Z., Wang, Z. & Xu, Y. Interpreting and Mitigating Hallucination in MLLMs through Multi-agent Debate. *arXiv preprint arXiv:2407.20505* (2024).
28. Wei, J., *et al.* Chain-of-thought prompting elicits reasoning in large language models. *Advances in neural information processing systems* **35**, 24824-24837 (2022).
29. Slonim, N., *et al.* An autonomous debating system. *Nature* **591**, 379-384 (2021).
30. Firth, H.V., *et al.* DECIPHER: Database of Chromosomal Imbalance and Phenotype in Humans Using Ensembl Resources. *Am J Hum Genet* **84**, 524-533 (2009).
31. Zhang, S. *Compendium of China's first list of rare diseases*, (People's Medical Publishing House, 2018).
32. Zhou, W., *et al.* TransVar: a multilevel variant annotator for precision genomics. *Nature Methods* **12**, 1002-1003 (2015).
33. Yang, J., *et al.* Enhancing phenotype recognition in clinical notes using large language models: PhenoBCBERT and PhenoGPT. *Patterns* **5**, 100887 (2024).
34. Wang, K., Li, M. & Hakonarson, H. ANNOVAR: functional annotation of genetic variants from high-throughput sequencing data. *Nucleic acids research* **38**, e164-e164 (2010).

35. Cingolani, P., *et al.* A program for annotating and predicting the effects of single nucleotide polymorphisms, SnpEff: SNPs in the genome of Drosophila melanogaster strain w1118; iso-2; iso-3. *Fly (Austin)* **6**, 80-92 (2012).
36. Stenson, P.D., *et al.* Human Gene Mutation Database (HGMD): 2003 update. *Hum Mutat* **21**, 577-581 (2003).
37. Landrum, M.J., *et al.* ClinVar: public archive of relationships among sequence variation and human phenotype. *Nucleic Acids Res* **42**, D980-985 (2014).
38. Steinhaus, R., *et al.* MutationTaster2021. *Nucleic Acids Research* **49**, W446-W451 (2021).
39. Ng, P.C. & Henikoff, S. SIFT: Predicting amino acid changes that affect protein function. *Nucleic acids research* **31**, 3812-3814 (2003).
40. Adzhubei, I.A., *et al.* A method and server for predicting damaging missense mutations. *Nat Methods* **7**, 248-249 (2010).